\documentclass[aps,prl,twocolumn]{revtex4}
\usepackage{graphicx}
\usepackage{amssymb}

\def\d{{\rm d}}

\def\eqref#1{(\ref{#1})}
\def\PRE{{\it Phys. Rev. E} }
\def\JSP{{\it J. Stat. Phys.} }
\def\EPJB{{\it Euro. Phys. J. B} }
\def\JPA{{\it J. Phys. A: Math. Gen.} }
\def\PRL{{\it Phys. Rev. Lett.} }
\def\av#1{\langle#1\rangle}
\renewcommand{\u}{{\bar u}}
\renewcommand{\v}{{\bar v}}

\begin{document}
\title{Condensation and coexistence in a two-species driven model}

\author{C. Godr\`eche$^{(a)}$, E. Levine$^{(b)}$, and D. Mukamel$^{(b)}$ }
\affiliation{$^{(a)}$ Service de Physique de l'\'Etat Condens\'e,
CEA Saclay, 91191 Gif-sur-Yvette cedex, France
\\$^{(b)}$ Department of Physics of Complex Systems, Weizmann
Institute of Science, Rehovot, Israel 76100.}

\begin{abstract}
Condensation transition in two-species driven systems in a ring
geometry is studied in the case where current-density relation of
a domain of particles exhibits two degenerate maxima. It is found
that the two maximal current phases coexist both in the
fluctuating domains of the fluid and in the condensate, when it
exists. This has a profound effect on the steady state properties
of the model. In particular, phase separation becomes more
favorable, as compared with the case of a~single maximum in the
current-density relation. Moreover, a selection mechanism imposes
equal currents flowing out of the condensate, resulting in a
neutral fluid even when the total number of particles of the two
species are not equal. In this case the particle imbalance shows
up only in the condensate.
\end{abstract}
\maketitle


Many properties of the phase diagram of driven systems are known
to be determined by some overall features of the current-density
relation. For example, this relation serves as a starting point
for analyzing models of vehicular traffic \cite{Traffic}, where it
is termed the fundamental diagram. It is also a useful tool for
analyzing boundary induced phase transitions in one-dimensional
systems \cite{Krug91,Hager01}, and stability of shocks
\cite{Shocks}. The aim of this paper is to investigate
how these global features affect
the properties of condensation transitions in driven diffusive systems
(DDS) on a ring. To this end we analyze in detail the case where
the current-density relation has two degenerate maxima. This is
found to have far-reaching consequences on the emergence of phase
separation. It results in new features which are not present in
the previously studied case of a current-density relation with a
single maximum \cite{Kafri02,Kafri03,Evans04}.

Condensation transitions in one-dimensional DDS
have been studied in detail in recent years
\cite{Reviews}. In particular, it was suggested that on a
mesoscopic level one can describe the dynamics of a broad class of
two-species DDS by a zero-range process (ZRP)
\cite{Kafri02,Evans04}. In this description one views the
microscopic configuration of the model as a sequence of particle
domains, bounded by vacancies. Each domain is defined as a stretch
of particles of both types. Neighboring domains exchange particles
through their currents. The existence of condensation in these
models, analogous to Bose-Einstein condensation (BEC), was found
to be related to the dependence of these currents on the length of
the domains. A quantitative criterion for the existence of a
condensation transition in a ring geometry was thus suggested.
According to this criterion, if the asymptotic form of the current
for large domains of length $n$ behaves as $j_n \sim
j_\infty(1+b/n)$, with $b>2$, a condensation transition takes
place at a sufficiently high overall particle density. This form of the
current implies that, at criticality, the steady-state domain-size
distribution scales as $p_n \sim n^{-b}$ for large $n$.
As in the BEC the condensed phase is composed of a critical fluid
of fluctuating domains coexisting with a single macroscopically
large condensate.

The criterion has previously been applied to models where the
current-density curve $j_{\infty}(\eta)$ in the bulk of a domain
exhibits a single maximum \cite{Kafri02,Kafri03,Evans04}. Here we
apply this approach to a model where $j_{\infty}(\eta)$ has two
degenerate maxima, and examine the condensation transition in
cases where, on average, the density within the domains
lies between the two extremal values corresponding to the two maxima.
A simple physical picture for the
dynamics inside a domain is inferred from numerical simulations of the model.
This picture is substantiated
by analyzing the properties of a two-species ZRP for
modelling the collective dynamics of domains. Our main findings
are: (i) The density in each particle domain (whether a fluid or a
condensate) is not homogeneous. The two maximal current phases
coexist within each domain, with a sharp interface separating the
two. The density profile of each of these phases is
algebraic, as expected for maximal current phases. (ii) The
non-homogeneous density profile affects the finite size correction
of the current, leading to a finite-size correction coefficient
$B$ which is larger than the expected $b$ of homogeneous domains. For
example, when the number of particles of both species are equal, we
find $B \gtrsim 2b$. This makes phase separation in this model
more favorable. Exact solution of the ZRP in mean-field geometry
and numerical simulations of the one-dimensional ZRP
support this finding. (iii) In the condensed phase the fluid
domains are neutral, even in systems with non-equal number of
particles of the two species, leaving the condensate as the only
imbalanced domain. This is in contrast with the case in which the
current-density curve has a single maximum, where all domains,
fluid and condensate, have the same average density.

We now define the model. Consider a one-dimensional ring with $L$
sites. Each site $i$ can be either vacant ($0$) or occupied by a
positive ($+$) or a negative ($-$) particle. Positive particles
are driven to the right while negative particles are driven to the
left. In addition to the hard-core repulsion, particles are
subjected to short-range interactions through a potential
\begin{equation}
V=-\frac\epsilon4 \sum_i{s_is_{i+1}}\;, \label{eq:H}
\end{equation}
where $s_i=+1$ ($-1$) if site $i$ is occupied by a $+$ ($-$)
particle, and $s_i=0$ if site $i$ is vacant. The interaction
parameter $\epsilon$ satisfies $-1 < \epsilon < 1$ to insure
positive transition rates. The evolution of the model is defined
by a random-sequential local dynamics, whereby a pair of
nearest-neighbor sites is selected at random, and particles are
exchanged with the following rates:
\begin{equation}
\begin{array}{cccl}
+- &\to& -+ &\qquad\mbox{with rate } 1+\Delta V \\
+\,0 &\to& 0\,+ & \qquad\mbox{with rate } 1 \\
0\,- &\to& -\,0 & \qquad\mbox{with rate } 1 \;.
\end{array}
\label{eq:rates}
\end{equation}
Here $\Delta V$ is the difference in the potential $V$ between the
final and the initial configurations. This dynamics conserves the
number of particles of each species, $N_+$ and $N_-$,
or, equivalently, the overall particle
densities  in the system,  $\rho_\pm=N_\pm/L$.
For a given domain, i.e., a sequence of positive and negative charges
confined by vacancies, the relative density $\eta$ is simply the
fraction of positive particles in that domain.
This is a fluctuating quantity, both in time and from domain to domain.
Model \eqref{eq:rates} was studied on a ring geometry for positive
$\epsilon$ in \cite{Kafri03,Evans04}. Here we focus on the
negative $\epsilon$ region, where the current-density relation
exhibits two degenerate maxima, and study mainly the case $\rho_+=\rho_-$. The
non-equal density case is briefly considered at the end of this
Letter, and is studied in detail in \cite{prep}.

\begin{figure}[t]
 \centerline{\includegraphics[width=4truecm]{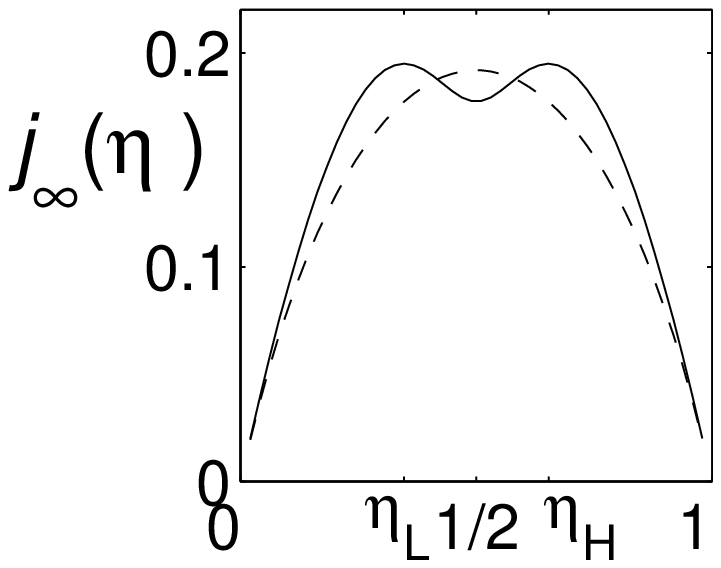}\hspace{0.4cm}\includegraphics[width=4.5truecm,viewport=0 -60 575 0]{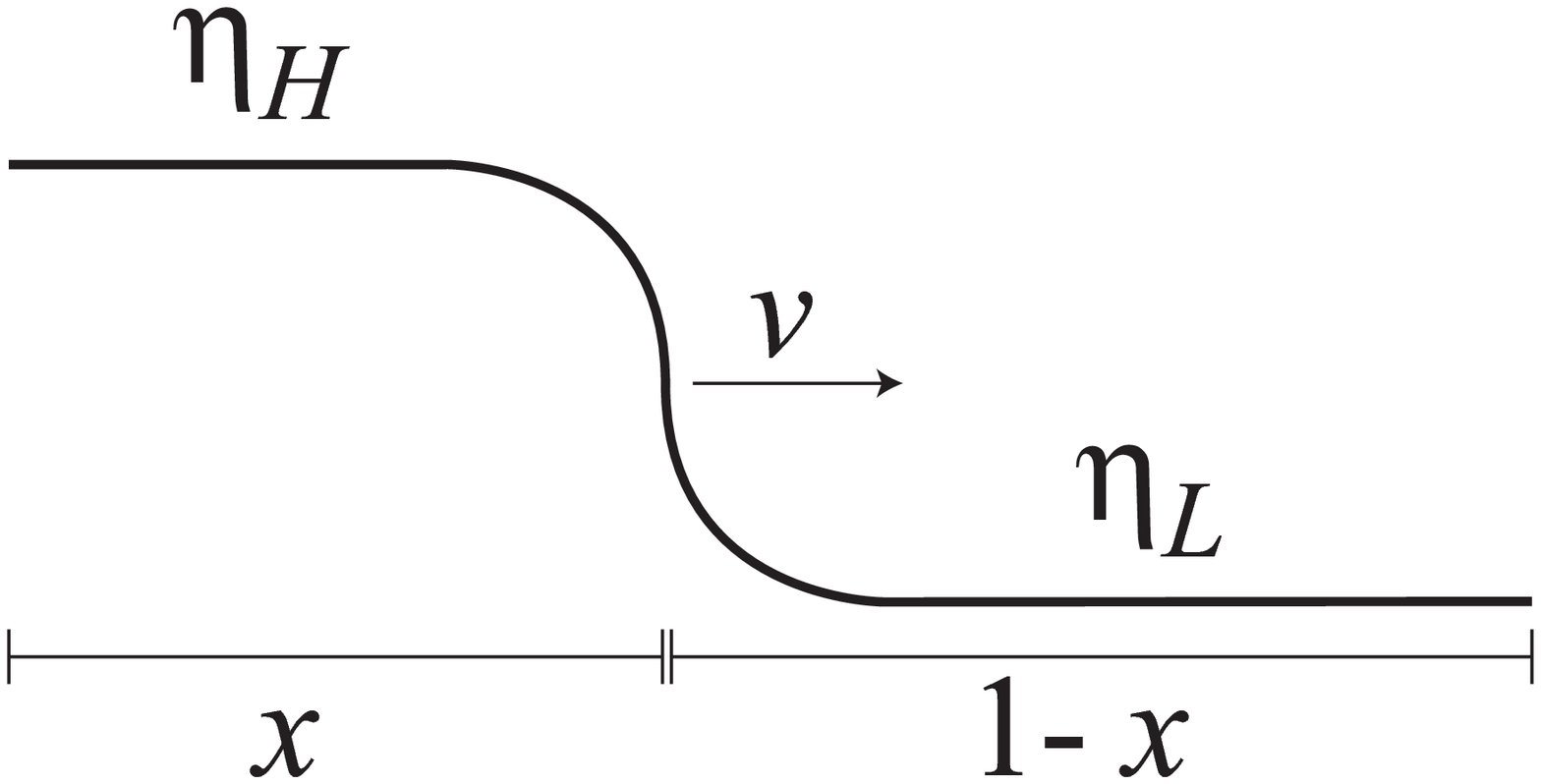}}
 \caption{(a) The current-density relation $j_\infty(\epsilon,\eta)$ for $\epsilon=0.5$
 (dashed line) and $\epsilon=-0.9$ (solid line). (b) A schematic picture of a typical snapshot of the density profile within a particle domain.}
 \label{fig:jofrho}
\end{figure}

\begin{figure}[t]
 \centerline{\includegraphics[width=8truecm]{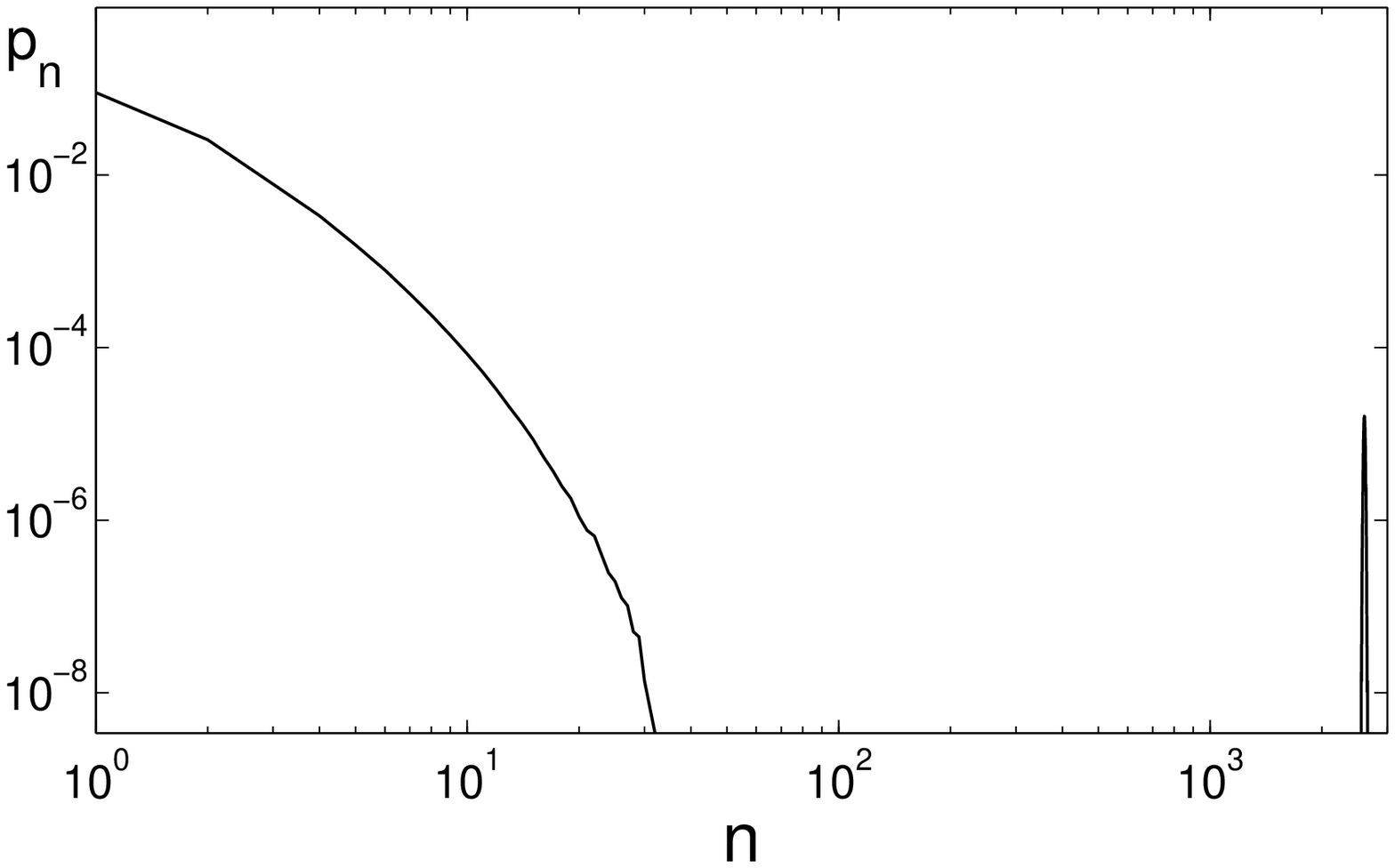}\hspace{-5.34cm}\includegraphics[width=4.3truecm,viewport=0   -375   538   0]{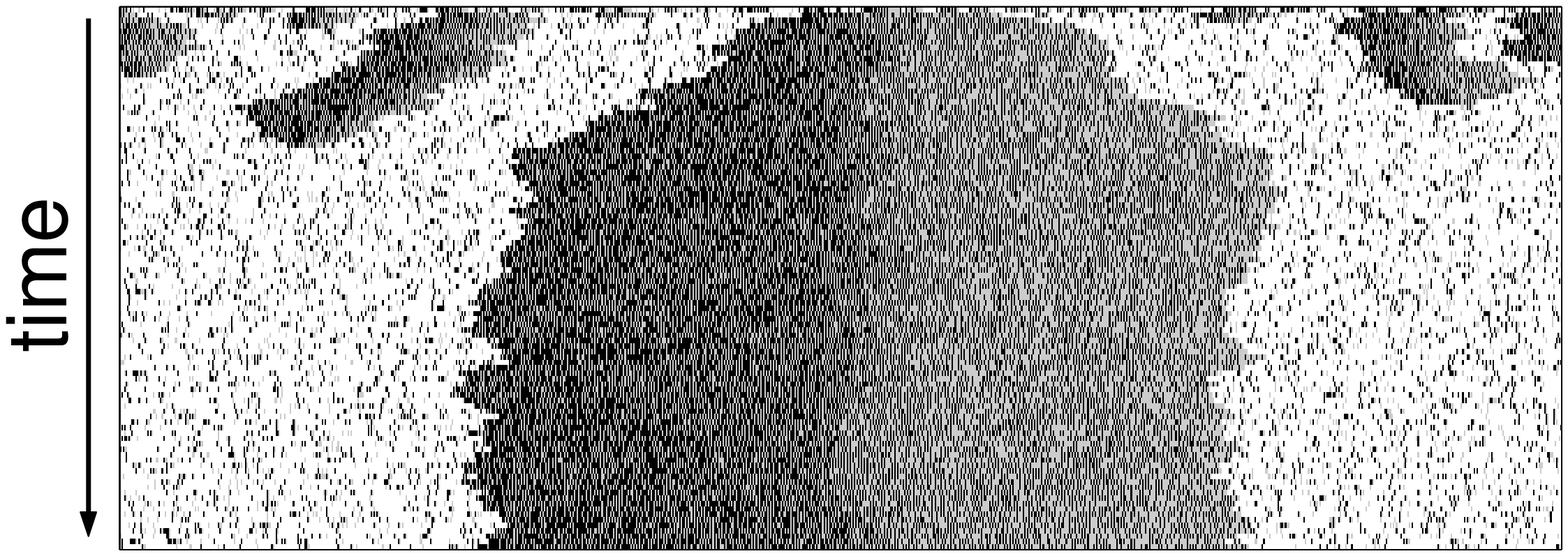}}
 \caption{Domain size distribution for the case $\epsilon=-0.9$. Simulation
 was performed on a system of size $L=5000$ with $N_+=N_-=1500$. Configurations of the model during coarsening towards the steady-state are presented in the inset. Positive particles are
 marked in black, negative particles in white, vacancies in gray. A configuration is presented every 500 Monte-Carlo
 sweeps. The system size is $L=1000$, $N_+=N_-=300$, and $\epsilon=-0.9$.}
 \label{fig:config}
\end{figure}

The region $\epsilon<0$ was studied in \cite{Hager01} in
open systems with the purpose of analyzing boundary-induced phase transitions. The
current-density relation of a domain of particles,
$j_\infty(\epsilon,\eta)$, was found to display a
single maximum at $\eta=1/2$ for $\epsilon\geq -0.8$, and two
degenerate maxima at $\eta_{H,L} = \frac12\left\{1 \pm [3-2
((\epsilon-1)/\epsilon)^{1/2}]^{1/2}\right\}$ for $\epsilon<-0.8$,
as depicted in Fig.~\ref{fig:jofrho}(a).

We carried out direct numerical simulations of the model for
$\epsilon<-0.8$. In Fig.~\ref{fig:config} we present the
domain-size distribution and typical configurations for
$\epsilon=-0.9$ at high densities $\rho_+=\rho_-$. This figure suggests the
existence of a pronounced macroscopic domain. Examining the
configurations it is evident that the relative density within the domains
is not homogeneous. Rather, it exhibits two coexisting regions,
corresponding to the two maximal-current phases. Indeed, the
densities of the two coexisting phases are equal to $\eta_{H,L}$,
and the current in the system is
$j_\infty(\eta_{H})=j_\infty(\eta_{L})$. This should be compared
with an open system driven with large boundary rates, where the
system assumes its maximal current and a similar coexistence takes
place \cite{Krug91}.

By itself, the appearance of a macroscopic domain in numerical
simulations of finite systems does not prove that condensation
takes place, as the presence of such a domain could result from a
finite size crossover \cite{Sharp}. The real question is whether
the macroscopic domain survives in the thermodynamic limit and
becomes a genuine condensate. To answer this question we use the
criterion for phase separation, and calculate the finite size
correction to the current of large domains. When the
current-density relation has a single maximum, the current of a
domain of length $n$ takes the asymptotic form $j_n \sim
j_\infty\left[1+b(\epsilon,\eta)/n\right]$, where
$b(\epsilon,\eta)$ is explicitly known~\cite{Evans04}, and where
in all domains $\eta$ is given by $N_+/(N_++N_-)$. In the present
case $b$ must be computed at the values of the density
corresponding to the two maxima of the current, $\eta=\eta_H$ or
$\eta=\eta_L$. For example we find
$b(\epsilon=-0.9,\eta=\eta_{H,L})\simeq1.14$. Applying the
criterion with this value of $b$ would then lead to the conclusion
that the existence of a macroscopic domain in
Fig.~~\ref{fig:config} is merely a finite-size effect. However, as
explained below, when a domain is composed of two coexisting
phases, the real finite size correction coefficient which enters
the expression of the current is not $b$, but an enhanced
coefficient $B\gtrsim 2b$, making phase separation more favorable.
In other words, $j_n \sim j_\infty\left[1+B(\epsilon)/n\right]$,
with $j_\infty=j_\infty(\epsilon,\eta=\eta_{H,L})$. In particular,
for $\epsilon=-0.9$ this yields $B>2$, implying that
Fig.~\ref{fig:config} corresponds to a genuine phase separation.

\begin{figure}[t]
 \centerline{\includegraphics[width=7truecm]{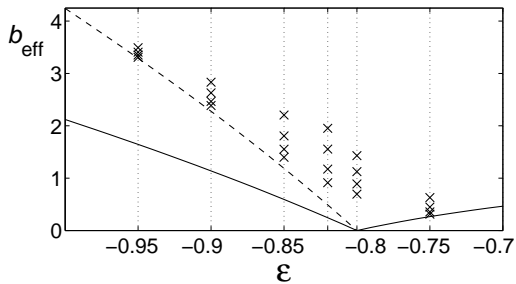}}
 \caption{The effective coefficient $b_{\rm eff}(\epsilon)$ measured in open systems of size $L=400, 800, 1600, 3200$ (from top to bottom).
 The lines correspond to  $b(\epsilon,\eta)$  (solid line) and $2b(\epsilon,\eta)$ (dashed line), with $\eta=1/2$ for $\epsilon>-0.8$ and $\eta=\eta_{H,L}$ for $\epsilon<-0.8$. }
 \label{fig:bofeopen}
\end{figure}

We first provide numerical evidence that indeed $B\gtrsim 2b$. It
is convenient to calculate the finite size correction to the current
$B(\epsilon)$ by simulating an isolated open domain of a fixed length
$n$ \cite{Kafri02}. The coefficient $B$ is then extracted by
measuring the effective coefficient at finite length $b_{\rm
eff}(n) = n\left(j_n/j_\infty-1\right)$ and extrapolating to $n
\to \infty$. In Fig.~\ref{fig:bofeopen} we present $b_{\rm
eff}(n)$ for various values of $\epsilon$ and system lengths. It
is found that, while for $\epsilon>-0.8$ the quantity $b_{\rm
eff}$ approaches $b(\epsilon,\eta=1/2)$ at large $n$, it is larger
than $b(\epsilon,\eta=\eta_{H,L})$ by a factor $\gtrsim 2$ for
$\epsilon<-0.8$. We note that higher order corrections become
significant as one approaches $\epsilon=-0.8$, where the leading
finite-size correction, $b(\epsilon,\eta=1/2)/n$, vanishes.

We now present a physical explanation of these observations. For a
domain of length $n$ each of the two coexisting phases occupies on
average only a length $n/2$. This effectively reduces the length
of the domain by a factor $2$, and thus the finite size correction
is expected to be about $\simeq 2b/n$ rather than $b/n$.
Quantifying this intuitive picture leads to an estimate of the
enhancement factor $A \equiv B(\epsilon)/b(\epsilon,\eta_{H,L})$.
We analyze the current emitted from a domain of length $n$ in the
fluid, composed of two coexisting maximal-current phases. A
schematic density profile in such a domain is given in
Fig.~\ref{fig:jofrho}(b). At the left side of the domain a
fraction $x$ of its length is occupied by a phase with high bulk
density $\eta_H$, while the remaining right side is occupied by
the other maximal-current phase, with low density $\eta_L$. The
position of the interface fluctuates in time around the midpoint,
i.e., on average, $\av{x}=1/2$. Numerical simulations strongly
suggest that the position $x$ varies on time scales which are much
longer than the equilibration time of the local density within
each phase \cite{prep}. We thus consider the dynamics of the
system on time scales which are short enough such that the
position of the interface $x$ and the size of the domain $n$ are
fixed. On these time scales the currents of, say, positive
particles $j_H(x)$ and $j_L(x)$ in the high and low density
phases, respectively, are given by
\begin{equation}
\label{eq:jljr} j_H(x) = j_\infty\left(1+\frac{b}{n\,x}\right),\;
j_L(x) = j_\infty\left(1+\frac{b}{n\,(1-x)}\right).
\end{equation}
Thus, as a result of the  flow of particles through the domain,
the interface moves with a velocity $v$, such that $j_H(x)-j_L(x)
= v\left(\eta_H-\eta_L\right). $ The outflow of particles from the
domain is therefore given by
\begin{equation}
\label{eq:outflow} j_H(x)-v\eta_H = j_L(x)-v\eta_L =
j_\infty\left[1+A(x)b/n \right]\;,
\end{equation}
where, using the expressions above, one has
\begin{equation}
\label{eq:A} A(x) =
\frac{1}{\eta_L-\eta_R}\left(\frac{\eta_L}{1-x}-\frac{\eta_R}{x}\right)\;.
\end{equation}
On longer time scales where the position of the interface $x$
fluctuates one has to average \eqref{eq:outflow} in order to
get the current emitted from the domain, leading to $A=
\av{A(x)}$. If the fluctuations in the position of the interface
do not scale with the domain size, then $\av{1/x} = 1/\av{x} = 2$,
and hence $A=2$. On the other hand, if these fluctuations scale
like the domain length, then $\av{1/x} > 1/\av{x}$, and $A>2$. In
the following we explore this question in more detail. Our
analysis suggests that indeed the width of the interface scales
with the domain length leading to $B > 2b$.

\begin{figure}[t]
 \centerline{\includegraphics[width=6.5truecm]{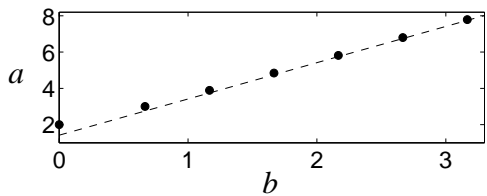}}
 \caption{The decay exponent $a$ for different values of $b$,
 as obtained from numerical integration of (\ref{eq:g}).
 The line is given by the large $b$ asymptotic form $2b+\sqrt{2}$. }
 \label{fig:aofb}
\end{figure}

Motivated by the discussion presented above, we introduce a
two-species ZRP which captures the main
features of the collective dynamics of the evolving domains. 
Consider a ring of $M$ boxes, where box $i$ contains $n_i$
particles, $k_i$ of which are positive and $l_i$ are negative:
$n_i=k_i+l_i$. The dynamics of the model is such that a box $i$ is
chosen at random and a positive charge is moved to its right
neighboring box with rate $u_{k,l}$ and a negative charge moves to
its left neighboring box with rate $v_{k,l}$. In this model a box
represents a generic domain of the original DDS, and the rates $u$ and
$v$ correspond to the outflow of particles from this domain,
as found in \eqref{eq:outflow}.  We thus take $u_{k,l}=1+A(x)b/n$ and
$v_{k,l}=1+A(1-x)b/n$.
The variable $x$ relates to the relative density $\eta$ by $x \eta_L + (1-x)
\eta_R = \eta=k/n$.
In what follows we analyze for simplicity
the case $\eta_L=1$ and $\eta_R=0$, which yields
\begin{eqnarray}\label{eq:uk10}
    u_{k,l}=1+\frac{b}{l}\;,\qquad v_{k,l}=1+\frac{b}{k}\;.
\end{eqnarray}
With this choice of rates the steady state of the model is not a
product measure \cite{Hanney03}, implying that no explicit description
of the stationary state is known.
We first consider the model
in the mean-field geometry, where all sites are connected. We
denote by $f_{k,l}$ the probability for a site to be occupied by
$k$ positive particles and $l$ negative particles. In the
thermodynamic limit $f_{k,l}$ obeys the evolution equation
\begin{eqnarray}\label{eq:mft}
\frac{\d f_{k,l}(t)}{\d t} &=&
u_{k+1,l}\,f_{k+1,l}+v_{k,l+1}\,f_{k,l+1}\nonumber\\&+&\u\,f_{k-1,l}(1-\delta_{k,0})+\v\,f_{k,l-1}(1-\delta_{l,0})\\
&-&\left[u_{k,l}(1-\delta_{k,0})+v_{k,l}(1-\delta_{l,0})+\u+\v\right]f_{k,l}\;,\nonumber
\end{eqnarray}
where $\bar{u}=\sum_{k,l}{u_{k,l}f_{k,l}}$ and
$\bar{v}=\sum_{k,l}{v_{k,l}f_{k,l}}$ are the $(+)$ and $(-)$
currents, respectively. In the continuum limit the steady state
distribution satisfies the following equation at criticality
($\u=\v=1$)
\begin{equation}
\frac{\partial^2 f_{k,l}}{\partial k^2}+\frac{\partial^2
f_{k,l}}{\partial l^2} +b\left(\frac{1}{l}\frac{\partial
f_{k,l}}{\partial k}+ \frac{1}{k}\frac{\partial f_{k,l}}{\partial
l}\right) =0\;. \label{crit}
\end{equation}
Moving to polar coordinates, and assuming the scaling solution
$f(r,\theta)=r^{-a}g(\theta)$, we find an equation for the angular
function $g(\theta)$
\begin{equation}\label{eq:g} \frac{\d^2
g(\theta)}{\d \theta^2} + \left( a-\frac{2b}{\sin
2\theta}\right)a\,g(\theta) =0\;,
\end{equation}
with the boundary conditions $g(0)=g(\pi/2)=0$. The determination
of the decay exponent $a$ is obtained by imposing the boundary
conditions. This is the quantization condition for this
Schr\"odinger equation. Except for special values of $b$ where $a$
can be determined exactly (e.g. $a=3$ for $b=2/3$), the value of $a$
as a function of $b$ is obtained by integrating~\eqref{eq:g}
numerically (Fig.~\ref{fig:aofb}). The large $b$ asymptotic form
obtained by the WKB approximation, $a\simeq 2b+\sqrt{2}$, agrees
very well with these results down to small $b$. From the predicted
form of the solution $f(r,\theta)$ we deduce that the domain size
distribution $p_n$, with $n=k+l$, scales as $p_n \sim n^{-(a-1)}$.
On the other hand, the rate out of a domain of size $n$, $j_n
\equiv \av{u_{k,l}}_{k+l=n}$, is of the form $j_n \sim 1+B/n$, and
we conclude that $B=a-1$.
Numerical integration of the temporal eqs.~(\ref{eq:mft}) gives
a decay exponent $a$ in perfect agreement with the predicted value
of the continuum limit.
This analysis shows that  $B\gtrsim 2b$, supporting the physical picture presented above.
This calculation was carried out within the
mean-field geometry and should not yield the exact values of
$B$ of the one-dimensional model. However, numerical simulations
of the latter indicate that $B$ is well
approximated by the mean-field result \cite{prep}. Coming back to
the DDS, the results above suggest that the position of the interface
inside a domain should
scale with the domain length.
This has been verified by numerical simulations \cite{prep}.

\begin{figure}[t]
 \centerline{\includegraphics[width=8truecm]{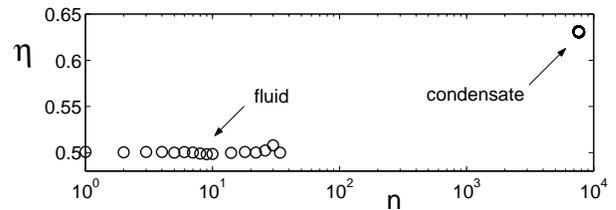}}
 \caption{The relative density $\eta$ of positive particles within a domain, for a system of size $10^4$ with $\epsilon=-0.9$. Here $N_+=5000$ and $N_-=3000$. }
 \label{fig:nonequal}
\end{figure}

So far we analyzed neutral system, where $\rho_+=\rho_-$. We now
consider the case of non-equal densities. While this case will be
studied in detail elsewhere \cite{prep}, here we only mention a
striking result: as long as $\eta_L< N_+/(N_++N_-) <
\eta_H$, all domains which reside in the fluid are equally
populated with positive and negative particles. The excess number
of particles of the majority species reside in the condensate.
This behavior is a result of the fact that the two currents
emitted from a domain of length $n$ are equal to the maximal
current, up to corrections of order $1/n$. The condensate is
therefore stationary in the thermodynamic limit even when the
densities are not equal. The condensate thus emits equal currents
of ($+$) and ($-$) particles. Hence domains in the fluid
cannot experience the fact that the overall densities in the
system are not equal.
Fig.~\ref{fig:nonequal} depicts the average
relative density $\eta$ of positive particles in domains of
various sizes, as measured in a large system for
$\epsilon=-0.9$.
It is readily seen that on average domains in the fluid are
neutral, whereas the relative density in the macroscopic domain
compensates for the excess of positive particles.

\begin{acknowledgments}
This work was partially carried out while CG was a Meyerhoff
Visiting Professor at the Weizmann Institute. Support of the
Albert Einstein Minerva Center for Theoretical Physics and the
Israel Science Foundation (ISF) is gratefully acknowledged.
\end{acknowledgments}

\end{document}